\documentclass{article}
\usepackage{spconf,amsmath,graphicx}
\usepackage{bm, amssymb, booktabs, multirow,cite, url, color}

\newcommand{\tabincell}[2]{\begin{tabular}{@{}#1@{}}#2\end{tabular}}

\title{DPCCN: DENSELY-CONNECTED PYRAMID COMPLEX CONVOLUTIONAL NETWORK FOR ROBUST SPEECH SEPARATION AND EXTRACTION}

\name{Jiangyu Han$^{1, 3}$, Yanhua Long$^{1,2}$\sthanks{Yanhua Long is the corresponding author. This work is supported by the National Natural Science Foundation of China (Grant No.62071302 and No.61701306), Czech National Science Foundation (GACR) project NEUREM3 (No.19-26934X), and Czech Ministry of Education, Youth and Sports (No. LTAIN19087).},
Luk\'{a}\v{s} Burget$^3$,  Jan \v{C}ernock\'{y}$^3$}
\address{
  $^1$Shanghai Normal University, Shanghai, China\\
  $^2$Unisound AI Technology Co., Ltd., Beijing, China\\
  $^3$Brno University of Technology, Faculty of Information Technology, Speech@FIT, Czechia}

\begin{document}
\ninept
\maketitle

\begin{abstract}
In recent years, a number of time-domain speech separation methods have been proposed.
However, most of them are very sensitive to the environments and wide domain coverage
tasks. In this paper, from the time-frequency domain perspective, we propose a
densely-connected pyramid complex convolutional network, termed DPCCN, to improve the robustness of speech separation under complicated conditions.
Furthermore, we generalize the DPCCN to target speech extraction (TSE) by integrating a new specially designed speaker encoder.
Moreover, we also investigate the robustness of DPCCN to unsupervised cross-domain TSE tasks. 
A Mixture-Remix approach is proposed to adapt the target domain acoustic characteristics for fine-tuning the source model. 
We evaluate the proposed methods not only under noisy and reverberant in-domain
condition, but also in clean but cross-domain conditions.
Results show that for both speech separation and extraction,
the DPCCN-based systems achieve significantly better performance and
robustness than the currently dominating time-domain methods,
especially for the cross-domain tasks. 
Particularly, we find that
the Mixture-Remix fine-tuning with DPCCN significantly outperforms
the TD-SpeakerBeam for unsupervised cross-domain TSE, with
around 3.5 dB SISNR improvement on target domain test set, 
without any source domain performance degradation.
\end{abstract}

\section{Introduction}
\label{sec:intro}

Speech separation (SS) aims to separate each source signal from 
mixed speech. Traditionally, it has been done in the time-frequency (T-F) domain \cite{dpcl, dan1, pit}. Recently, a convolutional
time-domain audio separation network (Conv-TasNet) \cite{tasnet} was proposed, and
has shown significant performance improvements over previous T-F
based techniques. Since then, several works have focused
on the time-domain methods, such as the DPRNN \cite{dprnn}, DPTNet \cite{dptnet},
SepFormer \cite{sepformer}, etc. Due to the success of speech
separation, researchers started to focus on the target speech extraction (TSE) --- 
a sub-task of speech separation requiring additional
target speaker clues to extract only a single speaker speech. 
Similarly, most of these works are also time-domain related, such as the
TD-SpeakerBeam \cite{tsb}, SpEx+ \cite{spex+} and channel-decorrelation \cite{cd2}.

In the current wave of time-domain based speech separation research,
we may however ask ``is it really the best to separate speech in time-domain?''
Reviewing the latest time-domain results on WSJ0-2mix \cite{dpcl},
a benchmark dataset of speech separation, the scale invariant
signal-to-noise ratio (SISNR) \cite{sisnr} has reached 22.3 dB \cite{sepformer}.
This is an excellent result and if we could achieve it on real speech separation, this SISNR value would be already good enough for real applications. However, authors in \cite{whamr} show that,
when evaluating the Conv-TasNet on a noisy and reverberated WSJ0-2mix dataset,
the separation performance is significantly reduced.
Besides, results in \cite{demystifying} also show that,
using time-frequency domain techniques can deliver
superior separation performance under complicated reverberant conditions.
This may due to the fact that, for the time-domain speech separation systems,
all speech transforms are learnt directly from the raw input waveforms.
These transforms may be very sensitive to the environment variations, especially
for the cross-domain tasks, in which the test conditions deviate
far from the conditions of training dataset.

\vspace{-0.04cm}
In addition, most state-of-the-art speech separation systems are supervised ones,
they are trained to separate the sources from
simulated mixtures created by adding up isolated ground-truth sources\cite{tasnet,dprnn}.
This reliance on ground-truth precludes scaling to widely available real-world mixture data
and limits the progress on open-domain tasks. Therefore, few latest
works start to focus on the unsupervised/semi-supervised
speech separation, such as \cite{hoshen2019towards},
mixture invariant training\cite{mixit} and its extensions
\cite{waspaa21,zhang2021teacher}. How to well exploit the real-world
unsupervised mixtures to boost the current separation and extraction systems
becomes very fundamental, important, and challenging.

\vspace{-0.04cm}
In this paper, we focus on the noise and reverberation, cross-domain 
robustness for both speech separation and extraction
from the time-frequency perspective. A densely-connected pyramid complex
convolutional network (DPCCN) is first proposed. It is motivated
by the network architecture in \cite{denseunet} (we call it DenseUNet
for simplicity) -- a U-Net \cite{unet} based
structure that combines temporal convolutional network (TCN) \cite{tcn}
and DenseNet \cite{densenet} to enhance the separation ability. Our
improvements are: 1) the mixture magnitude is discarded, only its
complex spectrum is taken as input to the network; 2) the decoder outputs
are waveforms instead of the real and imaginary (RI) components
of separated speech spectrum, and we replace the RI related loss function to the
negative SISNR; 3) a pyramid pooling layer \cite{pyramid} is added
at the end of the decoder to exploit more discriminative global information.
Besides, we design a new speaker encoder to generalize the
DPCCN to extract only the target speech for TSE task.
Furthermore, to improve the system robustness for unsupervised cross-domain TSE, a novel and effective Mixture-Remix approach is proposed to adapt the trained source model to the target domain acoustic characteristics. 

%\vspace*{-0.1cm}
On the noisy and reverberant in-domain LibriSpeech dataset \cite{indomain}, the proposed
DPCCN achieves more than 1.4~dB absolute SISNR improvement
over all listed state-of-the-art time-domain speech separation methods.
For the cross-domain speech separation and extraction tasks, we evaluate the
proposed approaches on clean Libri2Mix \cite{libri2mix} and Aishell2Mix that
we created from Aishell-1 \cite{aishell} corpus.
Results show that the DPCCN-based systems are 
more robust and achieve significantly better performance than baselines.
In particular, on the unsupervised cross-domain TSE tasks,
our proposed Mixture-Remix with DPCCN fine-tuning significantly outperforms
the TD-SpeakerBeam \cite{tsb}, a 3.5~dB absolute improvement in SISNR is
obtained on target domain test set without any performance degradation on source domain.

\section{Proposed Methods}
\label{sec:proposed_methods}

\begin{figure}[t]
  \centering
  \setlength{\abovecaptionskip}{0.cm}
  \includegraphics[height=7.0cm, width=8.6cm]{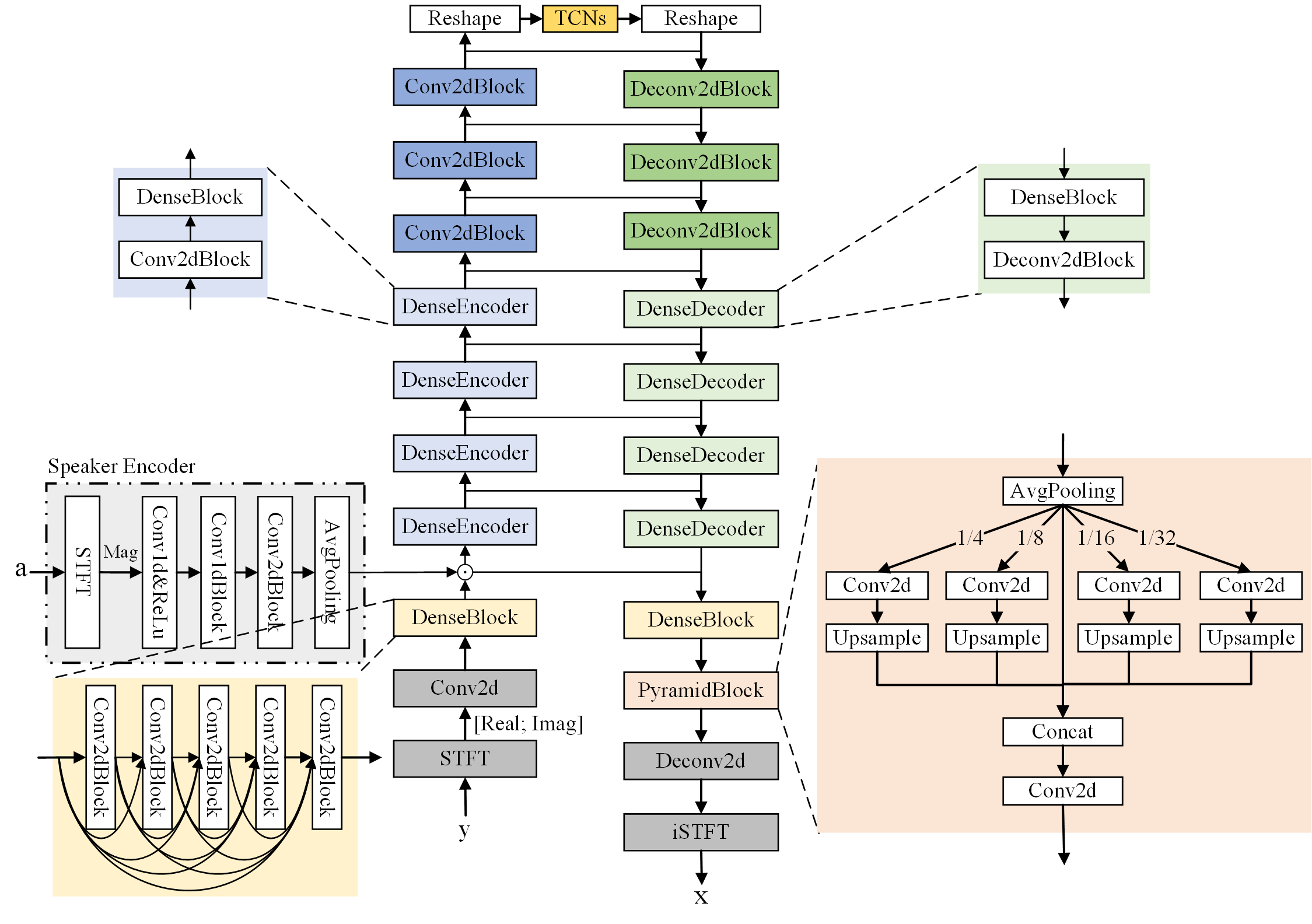}
  \caption{Structure of DPCCN with speaker encoder (sDPCCN) for target speech extraction.}
  \label{fig:dpccn}
  \vspace{-0.3cm}
\end{figure}

\subsection{DPCCN}
\label{ssec: dpccn}

DPCCN follows a U-Net style to encode mixture spectrum into a high-level representation, then decodes it into the clean speech. In DPCCN, DenseNet is used to alleviate the vanishing-gradient problem and
encourage the feature reuse; TCN is clamped between the codec to leverage
long-range time information; A pyramid pooling layer is introduced after decoder
to improve its global modeling ability. This architecture is  used for speech separation. To generalize
it for target speech extraction, an additional speaker encoder
is designed to encode the target speaker information to guide the network in extraction of the target speech.
The whole structure of DPCCN with the speaker encoder
(termed sDPCCN) is shown in Fig.~\ref{fig:dpccn},
where \textbf{y} is the mixture, \textbf{x} is
estimated speech, and \textbf{a} is target speaker enrollment
speech for TSE task.

The design of DPCCN is inspired by DenseUNet \cite{denseunet}. We improve it in the following ways: First, we discard the
mixture magnitude and only use its real and imaginary (RI) parts
as inputs to the network. This is because the RI parts
inherently contain the magnitude information.
Second, the outputs of DPCCN are waveforms instead of the RI parts of separated speech spectrum.
Correspondingly, we replace the original RI related loss function
to the standard negative SISNR, which has been proved to be
effective for speech separation.
Third, we introduce a pyramid pooling layer at the end of the decoder, 
considering that the limited receptive field of convolutional
network may make the U-Net unable to sufficiently incorporate the useful
global information. In the pyramid layer, different levels of
global information are obtained by averaging the feature map
at different scales, followed by bilinear interpolation upsampling
and concatenation operations to form the final feature representation
that ultimately carries both local and global context information.

In addition, we design a special speaker encoder
to generalize the DPCCN for extracting only the target speech.
In the speaker encoder, the magnitude spectrum of enrollment speech
is first processed through a 1-D convolution followed by rectified
linear unit (ReLU) function. Then a standard 1-D convolution
block with channel-wise layer normalization \cite{tasnet}
is used to model the temporal information. Next, it output is reshaped into a 4-D tensor and processed by the
following 2-D convolution block. Finally, the average pooling operation
along the time axis is used to aggregate the information of
target speaker to guide DPCCN to extract the target speech
through element-by-element multiplication.

\subsection{Mixture-Remix Fine-tuning}
\label{ssec: mix-remix}

\begin{figure}[t]
  \centering
  \includegraphics[width=8.0cm]{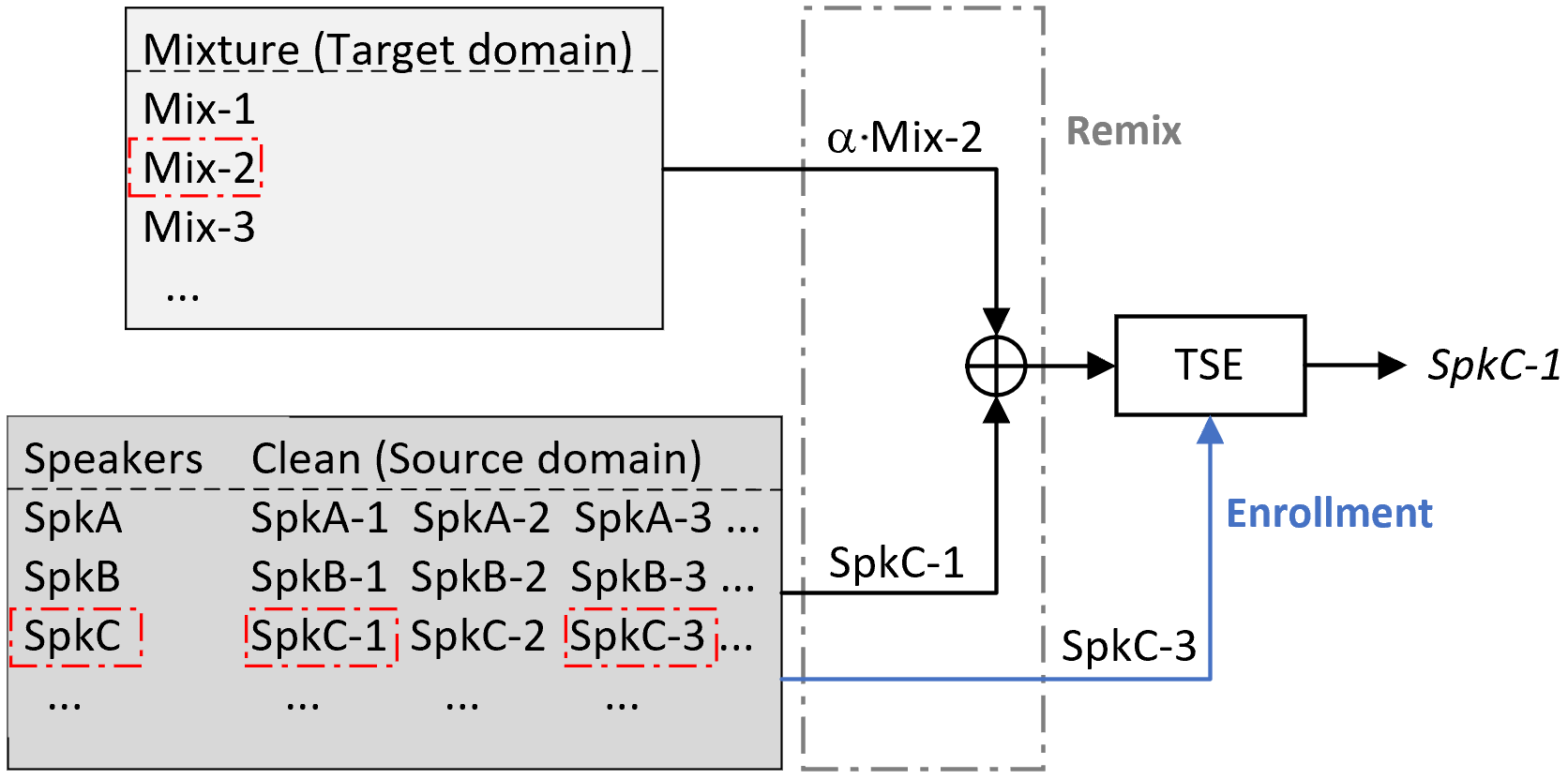}
  \caption{Illustration of Mixture-Remix fine-tuning. Red dashed box means random selection. $\alpha$ is a scaling factor converted from SNR between the selected mixture and the clean source.}
  \label{fig:mix_remix}
  \vspace{-0.3cm}
\end{figure}

The Mixture-Remix fine-tuning approach aims to improve the cross-domain robustness and performance of TSE systems under real scenarios, where only real-world target domain mixture data, and simulated source domain mixtures with isolated ground-truth source signals are available. We do not count on having any isolated ground-truth of target domain
training mixtures. We call this scenario as ``unsupervised cross-domain TSE'', and the target domain mixtures are ``unsupervised
mixtures''. Since most state-of-the-art TSE systems still heavily rely on the simulated training data, there is a huge performance gap
between in-domain and cross-domain test conditions, due to 
acoustic mismatch between the source and target domain signals.
Therefore, how to well utilize the available real mixtures to improve
the TSE system is a challenging open question.

In Fig.~\ref{fig:mix_remix}, an efficient and simple Mixture-Remix
approach is proposed. It generates new supervised TSE training data by remixing the real-world unsupervised mixtures of target domain
with the isolated sources of source domain. In this way, the acoustic characteristics of both target and source domain can be captured.
Then, the new supervised
training data is used to fine-tune the TSE model that already
well trained on the simulated source domain data. By doing so, the
cross-domain acoustic mismatch can be automatically alleviated.
The remixing is performed as illustrated in Fig.\ref{fig:mix_remix}:
given a random target domain unsupervised mixture \verb"Mix-2",
we randomly select a speaker \verb"SpkC" with utterance \verb"SpkC-1"
and \verb"SpkC-3" from the source domain.
Then, one of the selected utterance \verb"SpkC-1", is used to
remix with \verb"Mix-2" to generate a three-speakers mixture,
and \verb"SpkC-3" is taken as the target speaker enrollment speech
to the new three-speakers mixture for supervision to extract \verb"SpkC-1".
During the remixing, each unsupervised mixture is weighted by scaling factor $\alpha$
to produce more discriminative training mixtures as follows:
\begin{equation}
\label{eq:weight}
\alpha = \sqrt{\frac{\rm E_s}{{10^\frac{\rm SNR}{10}} \cdot \rm E_m}}, 
\end{equation}
where $\rm E_s$ and $\rm E_m$ are energy of the selected isolated source signal
and the unsupervised mixture belonging to the source and target domains, respectively.

\section{Task Construction}
\label{sec:task}

\textbf{In-domain SS task}: we use the noisy reverberant dataset
proposed in \cite{indomain}. This simulated dataset sampled at 16 kHz contains 20k,
5k and 3k utterances  in training, validation and test sets, respectively.
Two speakers from the 100-hour Librispeech \cite{librispeech}
and one noise from 100 Nonspeech Sounds \cite{100noise} are mixed
to generate each mixture. Each utterance is convolved with
room impulse responses generated by the image method \cite{rir} with
reverberation time ranging from 0.1s to 0.5s.
More details can be found in \cite{indomain}.

\textbf{Cross-domain SS and TSE tasks}: the English Libri2Mix \cite{libri2mix}
and Mandarin Aishell2Mix are used as the supervised source domain and unsupervised
target domain dataset, respectively.
Each mixture in Aishell2Mix is generated by mixing two speakers' utterances
from Aishell-1 \cite{aishell}. These utterances are randomly clamped
to 4 seconds and rescaled to
a random relative SNR between 0 and 5 dB.
This simulated Aishell2Mix is taken as the available real-world unsupervised
mixture data as described in section~\ref{ssec: mix-remix}.
For TSE task, the first speaker in the mixture of Libri2Mix is taken as the target speaker; 
its enrollment speech is randomly selected such that it differs from the one
in the mixture. While in the Aishell2Mix, only the test set is provided with the ground-truth speech, the others are all unsupervised mixtures.
We resample all the data to 8kHz. Details of the cross-domain tasks are shown in Table~\ref{tab:task}.
Both experimental datasets and their simulation code are
publicly available\footnote{\url{https://github.com/jyhan03/icassp22-dataset}}.

\vspace{-0.4cm}
\begin{table}[h]
  \caption{Setup of cross-domain SS and TSE tasks.
  ``Oracle" and ``Enroll" indicate whether the oracle (ground-truth)
  data and enrollment speech are available.}
  \label{tab:task}
 \setlength{\tabcolsep}{1.2mm}
  \centering
  \begin{tabular}{c|c|c|c|c|c|c}
    \hline
	Dataset & Type & \#Spks & \#Utts & Hours & Oracle & Enroll  \\
    \hline
    \multirow{3}{*}{\tabincell{c}{Libri2Mix\\(Source Domain)}} & train & 251 & 13900 & 58 & \checkmark & \checkmark \\
         & dev & 40 & 3000 & 11 & \checkmark & \checkmark \\
         & test & 40 & 3000 & 11 & \checkmark & \checkmark \\
    \hline
    \multirow{3}{*}{\tabincell{c}{Aishell2Mix\\(Target Domain)}}  & train & 340 & 10000 & 11 & - & - \\
        & dev & 40 & 3000 & 3 & - & - \\
        & test & 20 & 3000 & 3 & \checkmark & \checkmark \\
  	\hline
  \end{tabular}
  \vspace{-0.1cm}
\end{table}

\section{Experiments}
\label{sec:exp}

\subsection{Configurations}

Each \verb"Conv2dBlock" in DPCCN includes a 2-D convolution,
exponential linear units (ELU), and instance normalization (IN).
Each \verb"Deconv2dBlock" includes a 2-D deconvolution, ELU and IN.
The \verb"TCNs" contain 2 layers TCN, and each includes 10 TCN blocks 
composed by IN, ELU and 1-D convolution
with dilation factors $1, 2, 4, ..., 2^9$.
In \verb"PyramidBlock", all convolution kernel and stride sizes are set
to 1, the input/output channels of 2-D convolution before upsampling
and after concatenation are 32/8 and 64/32, respectively.
All other kernel and stride sizes, paddings, and convolution channels
are the same as the DenseUNet in \cite{denseunet}.

For (i)STFT layer in DPCCN, we use the square root of Hanning window with
FFT size of 512 and hop size of 128. Global mean-variance normalization
is applied to all input features. We train all models with Adam \cite{adam}
optimizer, 100 epochs on 4-second long segments, and the initial
learning rate is set to 0.001. Negative SISNR loss function is
used for both SS and TSE system training.
Utterance-level permutation invariant training (uPIT) \cite{pit}
is used for SS to address the source permutation problem.
SISNR\cite{sisnr} and its improvement, SISNRi, are used to evaluate the performance.
All fine-tuning related experiments are finished
within 20 epochs.

% The initial learning rate is set to 0.001, which is halved if the accuracy of validation set is not improved in 3 consecutive epochs.
% Early stopping is applied if no improvements in the validation set for 6 consecutive epochs.
% Adam \cite{adam} is used as the optimizer.
% Whether it is SS or TSE, we always use negative SISNR as loss function.

\subsection{Results and discussion}
\subsubsection{In-domain task}

We first compare  robustness of
our proposed DPCCN to noise and reverberation with several state-of-the-art separation methods
for standard in-domain speech separation in
Table~\ref{tab:in_domain}, where the DPRNN, DPTNet and
Sudo rm -rf results are the same as reported in \cite{g3c}, and the results
of Conv-TasNet, DCCRN and DenseUNet are produced by ourselves on the
same dataset. It is clear that our proposed
DPCCN significantly outperforms all other systems. Particulary,
the DPCCN achieves a 1.4 dB absolute SISNR improvement over the
best time-domain system DPRNN\cite{dprnn}. Such a performance
gain indicates that the DPCCN is a good candidate for
robust speech separation under complicated conditions.
In addition, adding the pyramid layer can bring
0.5 dB SISNR improvement, and the magnitude spectrum does not
bring any benefits. We also trained the DPCCN with the loss function reported
in DenseUNet\cite{denseunet}, and we found that using the negative SISNR
loss function can bring a 1.1 dB improvement in SISNR.

\begin{table}[t]
  \caption{SISNR/SISNRi performance  of different speech separation systems on the noisy and reverberant mixed LibriSpeech dataset.}
  \label{tab:in_domain}
  \centering
  \begin{tabular}{l|c|c}
    \hline
	System & Domain & SISNR / SISNRi \\
    \hline
    Conv-TasNet \cite{tasnet} & Time & 8.3 / 8.72  \\
    DPRNN \cite{dprnn} & Time & 9.0 / 9.42  \\
    DPTNet \cite{dptnet}  & Time & 8.1 / 8.52  \\
    Sudo rm -rf \cite{sudo} & Time & 6.8 / 7.22  \\
    DCCRN \cite{dccrn}  & Frequency & 7.2 / 7.62  \\
    DenseUNet \cite{denseunet} & Frequency & 8.78 / 9.19  \\
    \hline
    Proposed DPCCN & Frequency & \textbf{10.40 / 10.82}  \\
     \qquad w/ magnitude spectrum & Frequency & 10.36 / 10.78 \\
     \qquad w/o pyramid layer & Frequency & 9.88 / 10.30 \\
    \hline
  \end{tabular}
   \vspace{-0.2cm}
\end{table}

\subsubsection{Cross-domain tasks}

Another way to verify the system robustness is to investigate its
performance behavior on cross-domain test sets. Table~\ref{tab:cross_domain}
compares the performance of our proposed DPCCN and sDPCCN with two strong
baselines, the Conv-TasNet \cite{tasnet} and the TD-SpeakerBeam \cite{tsb}
for SS and TSE tasks, respectively. We also calculate an ``ST-Gap" to show the
cross-domain performance gap more clearly. It is defined as the SISNR difference
between source and target domain test sets divided by the source domain SISNR,
the lower the better.

\begin{table}[t]
  \caption{SISNR/SISNRi performance of speech separation and target speech extraction on Libri2Mix and Aishell2Mix. ``TSB" means TD-SpeakerBeam.
  ``ST-Gap'' means the relative SISNR degradation between the source domain and the target domain, the lower the better.
  Systems are all trained on Libri2Mix.}
  \label{tab:cross_domain}
%	\small
  \centering
  \setlength{\tabcolsep}{1.9mm}
  \begin{tabular}{l|c|c|c|c}
    \hline
	Task & System & Libri2Mix & Aishell2Mix & ST-Gap\\
    \hline
    \multirow{2}{*}{SS} & Conv-TasNet & 11.98 / 11.98 & 2.08 / 2.08 & $\downarrow 82.6\%$ \\
        & DPCCN & 13.05 / 13.04 & 5.09 / 5.09 & $\downarrow 61.0\%$ \\
    \hline
    \multirow{2}{*}{TSE} & TSB & 12.20 / 12.23 & 1.85 / -0.65  & $\downarrow 84.8\%$ \\
        & sDPCCN & 11.57 / 11.61 & 5.78 / 3.28 & $\downarrow 50.0\%$ \\
  	\hline
  \end{tabular}
  \vspace{-0.3cm}
\end{table}

In Table \ref{tab:cross_domain}, all systems are trained on the source domain
Libri2Mix but tested on both in-domain (Libri2Mix) and out-of-domain (Aishell2Mix)
evaluation sets. Three findings are observed: 1) huge cross-domain performance
gap exists in both SS and TSE tasks, either from the absolute SISNR numbers, or from the ST-Gap values; 2) DPCCN always shows much better speech separation
performance than Conv-TasNet under both in-domain and cross-domain conditions;
3) significant target domain performance improvement (3.9 dB SISNR) is
obtained by sDPCCN over TD-SpeakerBeam, even with slightly worse results
on in-domain test set.  All these findings confirm that the state-of-the-art
time-domain methods, Conv-TasNet and TD-SpeakerBeam, are very sensitive to
domain deviations, even for Libri2Mix and Aishell2Mix that are both clean speech but in different languages. 
Our proposed (s)DPCCN shows more robustness, although
the absolute SISNR in target domain is still relative low.

\subsubsection{Mixture-Remix fine-tuning}

Table \ref{tab:remix} presents the target domain performance
improvements using our proposed Mixture-Remix fine-tuning for target
speech extraction. The 1st line results are upper bound (oracle)
performance of sDPCCN trained from Aishell2Mix with ground-truth,
and they also demonstrate the large cross-domain performance gap 
as observed in Table \ref{tab:cross_domain}.

\vspace{-0.2cm}
\begin{table}[!htbp]
\caption{SISNR/SISNRi performance of Mixture-Remix fine-tuning on TSE tasks.
``A-L-3Mix" is the three-speakers mixture dataset generated by remixing Aishell2Mix and LibriSpeech. ``TSB" refers to the TD-SpeakerBeam.}
  \label{tab:remix}
  \setlength{\tabcolsep}{0.8mm}
  \centering
	\begin{tabular}{l|c|c|c|c}
		\hline
		\multirow{2}{*}{System} & \multirow{2}{*}{Train} & \multirow{2}{*}{\tabincell{c}{Fine-tune\\(A-L-3Mix)}} & \multicolumn{2}{c}{Evalation} \\
		\cline{4-5}
		& & &Libri2Mix & Aishell2Mix\\
		\hline
		\multirow{4}{*}{sDPCCN} & Aishell2Mix & - & 2.72 / 2.75 & 9.44 / 6.93 \\
		\cline{2-5}
                 & \multirow{3}{*}{Libri2Mix} & - & 11.57 / 11.61 & 5.78 / 3.28 \\ \cline{3-5}
            & & \checkmark & 0.55 / 0.58 & 4.16 / 1.66 \\ \cline{3-5}
            & & + Libri2Mix & \textbf{11.83} / \textbf{11.87} & \textbf{6.61} / \textbf{4.11} \\ \cline{3-5}
           % & & + Aishell2Mix & 7.70 / 7.74 & 10.94 / 8.44 \\
           \cline{1-5}
        \multirow{2}{*}{TSB} & \multirow{2}{*}{Libri2Mix} & - & 12.20 / 12.23 & 1.85 / -0.65 \\ \cline{3-5}
        & & + Libri2Mix & 12.56 / 12.59 & 3.13 / 0.63 \\
		\hline
	\end{tabular}
%\vspace{-0.2cm}
\end{table}

The 3rd line results are from the sDPCCN system first
trained on the source domain Libri2Mix training set, and then fine-tuned
using only the A-L-3Mix data generated by our Mixture-Remix approach.
We find that the SISNR on Libri2Mix is extremely low (only 0.55 dB),
which indicates that the English sDPCCN source model has been well
adapted to the target Mandarin Aishell2Mix acoustic environments
by the Mixture-Remix fine-tuning, and the adapted target model deviates significantly
from the source model. Meanwhile, it is interesting to see that on Aishell2Mix, the fine-tuned performance, 4.16 dB is also much worse than the original 5.78 dB, without any fine-tuning operations. This may be due to
the fact that, the remixed A-L-3Mix mixtures contain three speakers 
whose acoustic conditions for target speaker are mismatched with the two-speakers mixed training and evaluation sets. 
Such a phenomenon indicates that
the current speech extraction process is very sensitive to the number
of speakers in the mixtures.

Therefore, we choose to add the two-speakers mixed source
domain training data Libri2Mix, new generated A-L-3Mix mixtures, and their corresponding enrollment speech together to fine-tune the whole source model. The results are shown in the 4th line.  
To our surprise, both the in-domain
and target domain TSE performance are improved: for  sDPCCN
system, the SISNR is improved from 11.57 to 11.83 dB, and
5.78 to 6.61 dB on Libri2Mix and Aishell2Mix test sets, respectively.
And these improvements are consistent for the time-domain TD-SpeakerBeam system.
All results indicate that, without
sacrificing the performance
on source domain test sets, the proposed Mixture-Remix fine-tuning
not only can significantly improve the target domain performance for
unsupervised cross-domain TSE scenarios, but also generalizes well for time-domain systems. 

\begin{figure}[t]
  \centering
  \includegraphics[width=8.5cm]{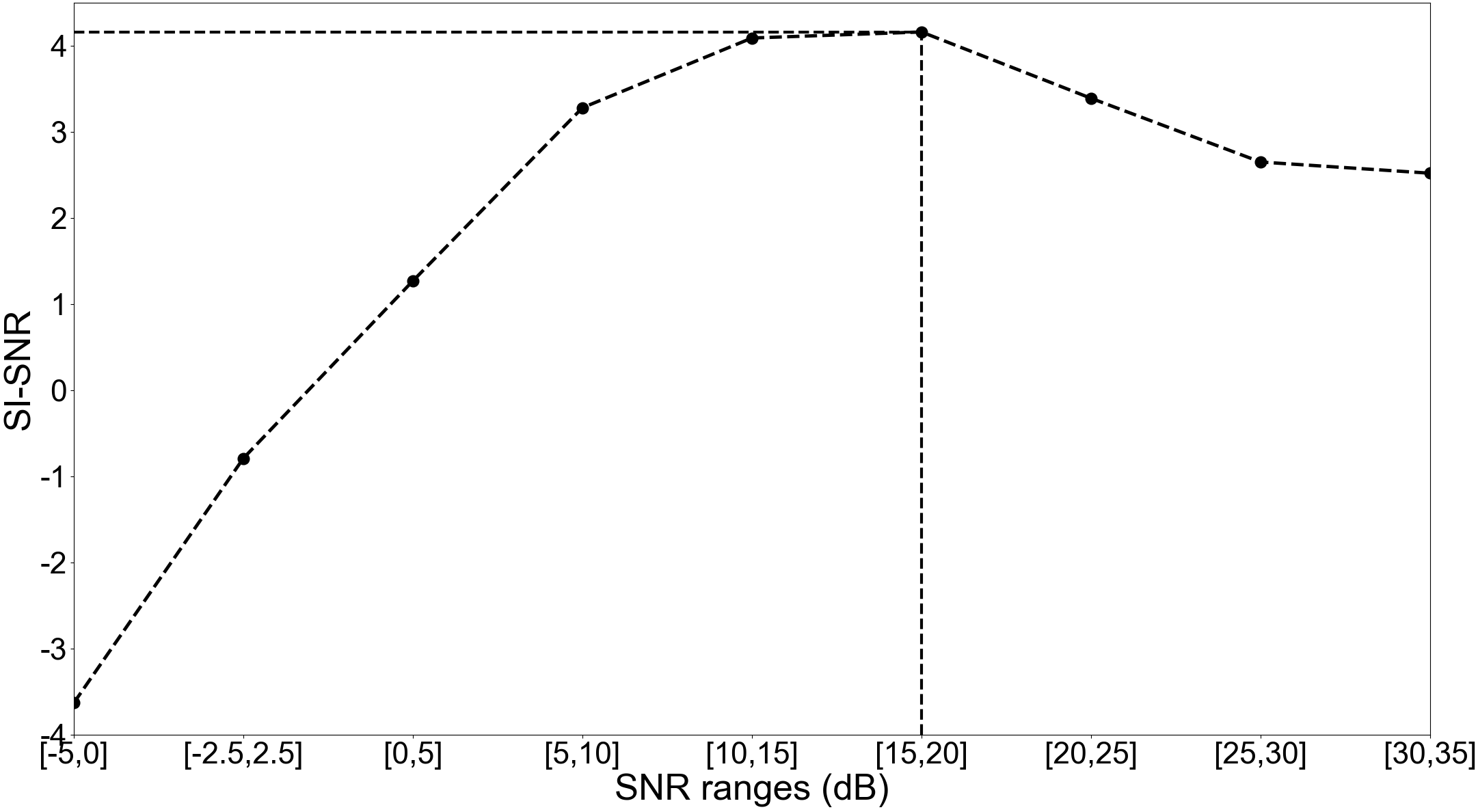}
  \caption{SISNR performance of sDPCCN system varying with different SNR ranges during remixing for Mixture-Remix fine-tuning.}
  \label{fig:snr_range}
  \vspace{-0.4cm}
\end{figure}

In addition, it is worth investigating how Mixture-Remix fine-tuning strategy
is influenced by the SNR range during remixing. Fig.\ref{fig:snr_range}
shows the SISNR performance on Aishell2Mix test set of sDPCCN system varying with different SNR ranges during the remixing stage. In the figure, the sDPCCN is only fine-tuned
using the A-L-3Mix data and we see that
a higher SNR range than that of the evaluation set (0 to 5 dB) is required to obtain better performance.
This may because it is more
difficult to extract the target speech from a three-speaker mixture
than from a two-speaker mixture. Therefore, for all the fine-tuning
experiments in Table \ref{tab:remix}, the best SNR range, from 15 to 20 dB is used.

\section{Conclusion}
\label{sec:conclusion}

In this study, we propose two novel methods for robust speech
separation and extraction. One is the densely-connected pyramid complex
convolutional network (DPCCN), and the other is a Mixture-Remix fine-tuning
approach. Their  robustness are  verified
not only in noisy and reverberant in-domain speech separation task, but also
examined in cross-domain speech separation and extraction tasks.
Experiments show that the proposed (s)DPCCN has much better performance
and robustness over other state-of-the-art separation and extraction
methods. Moreover, the Mixture-Remix fine-tuning is proved to be
effective to improve the target speech extraction system under
real-world unsupervised cross-domain scenarios. Our future work
will focus on generalizing the Mixture-Remix fine-tuning to different unsupervised speech separation tasks.

\bibliographystyle{IEEEbib}
\bibliography{refs}

\begin{thebibliography}{10}

\bibitem{dpcl}
J.~R. Hershey, Z.~Chen, J.~Le~Roux, and S.~Watanabe,
\newblock ``Deep clustering: Discriminative embeddings for segmentation and
  separation,''
\newblock in {\em Proc. ICASSP}, 2016, pp. 31--35.

\bibitem{dan1}
Z.~Chen, Y.~Luo, and N.~Mesgarani,
\newblock ``Deep attractor network for single-microphone speaker separation,''
\newblock in {\em Proc. ICASSP}, 2017, pp. 246--250.

\bibitem{pit}
M.~Kolb{\ae}k, D.~Yu, Z.~H. Tan, and J.~Jensen,
\newblock ``Multitalker speech separation with utterance-level permutation
  invariant training of deep recurrent neural networks,''
\newblock {\em IEEE/ACM Transactions on Audio, Speech, and Language
  Processing}, vol. 25, no. 10, pp. 1901--1913, 2017.

\bibitem{tasnet}
Y.~Luo and N.~Mesgarani,
\newblock ``{Conv-TasNet}: Surpassing ideal time--frequency magnitude masking
  for speech separation,''
\newblock {\em IEEE/ACM transactions on audio, speech, and language
  processing}, vol. 27, no. 8, pp. 1256--1266, 2019.

\bibitem{dprnn}
Y.~Luo, Z.~Chen, and T.~Yoshioka,
\newblock ``Dual-path {RNN}: efficient long sequence modeling for time-domain
  single-channel speech separation,''
\newblock in {\em Proc. ICASSP}, 2020, pp. 46--50.

\bibitem{dptnet}
J.~Chen, Q.~Mao, and D.~Liu,
\newblock ``Dual-path transformer network: Direct context-aware modeling for
  end-to-end monaural speech separation,''
\newblock in {\em Proc. Interspeech}, 2020, pp. 2642--2646.

\bibitem{sepformer}
C.~Subakan, M.~Ravanelli, S.~Cornell, M.~Bronzi, and J.~Zhong,
\newblock ``Attention is all you need in speech separation,''
\newblock in {\em Proc. ICASSP}, 2021, pp. 21--25.

\bibitem{tsb}
M.~Delcroix, T.~Ochiai, K.~Zmolikova, K.~Kinoshita, N.~Tawara, T.~Nakatani, and
  S.~Araki,
\newblock ``Improving speaker discrimination of target speech extraction with
  time-domain speakerbeam,''
\newblock in {\em Proc. ICASSP}, 2020, pp. 691--695.

\bibitem{spex+}
M.~Ge, C.~Xu, L.~Wang, E.~S. Chng, J.~Dang, and H.~Li,
\newblock ``{SpEx+}: A complete time domain speaker extraction network,''
\newblock in {\em Proc. Interspeech}, 2020, pp. 1406--1410.

\bibitem{cd2}
J.~Han, W.~Rao, Y.~Wang, and Y.~Long,
\newblock ``Improving channel decorrelation for multi-channel target speech
  extraction,''
\newblock in {\em Proc. Interspeech}, 2021, pp. 1847--1851.

\bibitem{sisnr}
J.~Le~Roux, S.~Wisdom, H.~Erdogan, and J.~R. Hershey,
\newblock ``{SDR}--half-baked or well done?,''
\newblock in {\em Proc. ICASSP}, 2019, pp. 626--630.

\bibitem{whamr}
M.~Maciejewski, G.~Wichern, E.~McQuinn, and J.~Le~Roux,
\newblock ``{WHAMR!}: Noisy and reverberant single-channel speech separation,''
\newblock in {\em Proc. ICASSP}, 2020, pp. 696--700.

\bibitem{demystifying}
J.~Heitkaemper, D.~Jakobeit, C.~Boeddeker, L.~Drude, and R.~Haeb-Umbach,
\newblock ``Demystifying {TasNet}: A dissecting approach,''
\newblock in {\em Proc. ICASSP}, 2020, pp. 6359--6363.

\bibitem{hoshen2019towards}
Y.~Hoshen,
\newblock ``Towards unsupervised single-channel blind source separation using
  adversarial pair unmix-and-remix,''
\newblock in {\em Proc. ICASSP}, 2019, pp. 3272--3276.

\bibitem{mixit}
S.~Wisdom, E.~Tzinis, H.~Erdogan, R.~J. Weiss, K.~Wilson, and J.~R. Hershey,
\newblock ``Unsupervised sound separation using mixture invariant training,''
\newblock in {\em Advances in Neural Information Processing Systems}, 2020.

\bibitem{waspaa21}
S.~Wisdom, A.~Jansen, R.~J. Weiss, H.~Erdogan, and J.~R. Hershey,
\newblock ``Sparse, efficient, and semantic mixture invariant training: Taming
  in-the-wild unsupervised sound separation,''
\newblock in {\em Workshop on Applications of Signal Processing to Audio and
  Acoustics}, 2021.

\bibitem{zhang2021teacher}
J.~Zhang, C.~Zorila, R.~Doddipatla, and J.~Barker,
\newblock ``{Teacher-student MixIT} for unsupervised and semi-supervised speech
  separation,''
\newblock in {\em Proc. Interspeech}, 2021, pp. 3495--3499.

\bibitem{denseunet}
Z.-Q. Wang and D.~Wang,
\newblock ``Count and separate: Incorporating speaker counting for continuous
  speaker separation,''
\newblock in {\em Proc. ICASSP}, 2021, pp. 11--15.

\bibitem{unet}
O.~Ronneberger, P.~Fischer, and T.~Brox,
\newblock ``{U-Net}: Convolutional networks for biomedical image
  segmentation,''
\newblock in {\em International Conference on Medical image computing and
  computer-assisted intervention}, 2015, pp. 234--241.

\bibitem{tcn}
S.~Bai, J.~Z. Kolter, and V.~Koltun,
\newblock ``An empirical evaluation of generic convolutional and recurrent
  networks for sequence modeling,''
\newblock {\em arXiv preprint arXiv:1803.01271}, 2018.

\bibitem{densenet}
G.~Huang, Z.~Liu, L.~Van Der~Maaten, and K.~Q. Weinberger,
\newblock ``Densely connected convolutional networks,''
\newblock in {\em Proc. CVPR}, 2017, pp. 4700--4708.

\bibitem{pyramid}
H.~Zhao, J.~Shi, X.~Qi, X.~Wang, and J.~Jia,
\newblock ``Pyramid scene parsing network,''
\newblock in {\em Proc. CVPR}, 2017, pp. 2881--2890.

\bibitem{indomain}
Y.~Luo, Z.~Chen, N.~Mesgarani, and T.~Yoshioka,
\newblock ``End-to-end microphone permutation and number invariant
  multi-channel speech separation,''
\newblock in {\em Proc. ICASSP}, 2020, pp. 6394--6398.

\bibitem{libri2mix}
J.~Cosentino, M.~Pariente, S.~Cornell, A.~Deleforge, and E.~Vincent,
\newblock ``{LibriMix}: An open-source dataset for generalizable speech
  separation,''
\newblock {\em arXiv preprint arXiv:2005.11262}, 2020.

\bibitem{aishell}
H.~Bu, J.~Du, X.~Na, B.~Wu, and H.~Zheng,
\newblock ``{AISHELL-1}: An open-source mandarin speech corpus and a speech
  recognition baseline,''
\newblock in {\em 2017 20th Conference of the Oriental Chapter of the
  International Coordinating Committee on Speech Databases and Speech I/O
  Systems and Assessment (O-COCOSDA)}, 2017, pp. 1--5.

\bibitem{librispeech}
V.~Panayotov, G.~Chen, D.~Povey, and S.~Khudanpur,
\newblock ``{LibriSpeech}: an asr corpus based on public domain audio books,''
\newblock in {\em Proc. ICASSP}, 2015, pp. 5206--5210.

\bibitem{100noise}
G.~Hu and D.~Wang,
\newblock ``A tandem algorithm for pitch estimation and voiced speech
  segregation,''
\newblock {\em IEEE/ACM transactions on audio, speech, and language
  processing}, vol. 18, no. 8, pp. 2067--2079, 2010.

\bibitem{rir}
J.~B. Allen and D.~A. Berkley,
\newblock ``Image method for efficiently simulating small-room acoustics,''
\newblock {\em The Journal of the Acoustical Society of America}, vol. 65, no.
  4, pp. 943--950, 1979.

\bibitem{adam}
D.~P. Kingma and J.~Ba,
\newblock ``Adam: A method for stochastic optimization,''
\newblock {\em arXiv preprint arXiv:1412.6980}, 2014.

\bibitem{g3c}
Y.~Luo, C.~Han, and N.~Mesgarani,
\newblock ``Group communication with context codec for lightweight source
  separation,''
\newblock {\em IEEE/ACM Transactions on Audio, Speech, and Language
  Processing}, vol. 29, pp. 1752--1761, 2021.

\bibitem{sudo}
E.~Tzinis, Z.~Wang, and P.~Smaragdis,
\newblock ``Sudo rm-rf: Efficient networks for universal audio source
  separation,''
\newblock in {\em 2020 IEEE 30th International Workshop on Machine Learning for
  Signal Processing (MLSP)}, 2020, pp. 1--6.

\bibitem{dccrn}
Y.~Hu, Y.~Liu, S.~Lv, M.~Xing, S.~Zhang, Y.~Fu, J.~Wu, B.~Zhang, and L.~Xie,
\newblock ``{DCCRN}: Deep complex convolution recurrent network for phase-aware
  speech enhancement,''
\newblock in {\em Proc. Interspeech}, 2020, pp. 2472--2476.

\end{thebibliography}
\end{document}